\documentclass[]{article}

\usepackage{color}
\usepackage{amsmath,amssymb}
\usepackage[usenames,dvipsnames,svgnames,table]{xcolor}
\usepackage{graphicx}

\small

\begin{document}

\twocolumn

\begin{center}
\fboxrule0.02cm
\fboxsep0.4cm
\fcolorbox{blue}{AliceBlue}{\rule[-0.9cm]{0.0cm}{1.8cm}{\parbox{7.8cm}
{ \begin{center}


{\large\bf GW Orionis:\vspace{0.1cm}\\ A pre-main-sequence triple with a warped disk and a torn-apart ring as benchmark for disk hydrodynamics}

\vspace{0.2cm}

{\large\em Stefan Kraus, University of Exeter\\}
{\large\em (email: s.kraus@exeter.ac.uk)}

\vspace{0.5cm}


\includegraphics[height=0.25\textwidth]{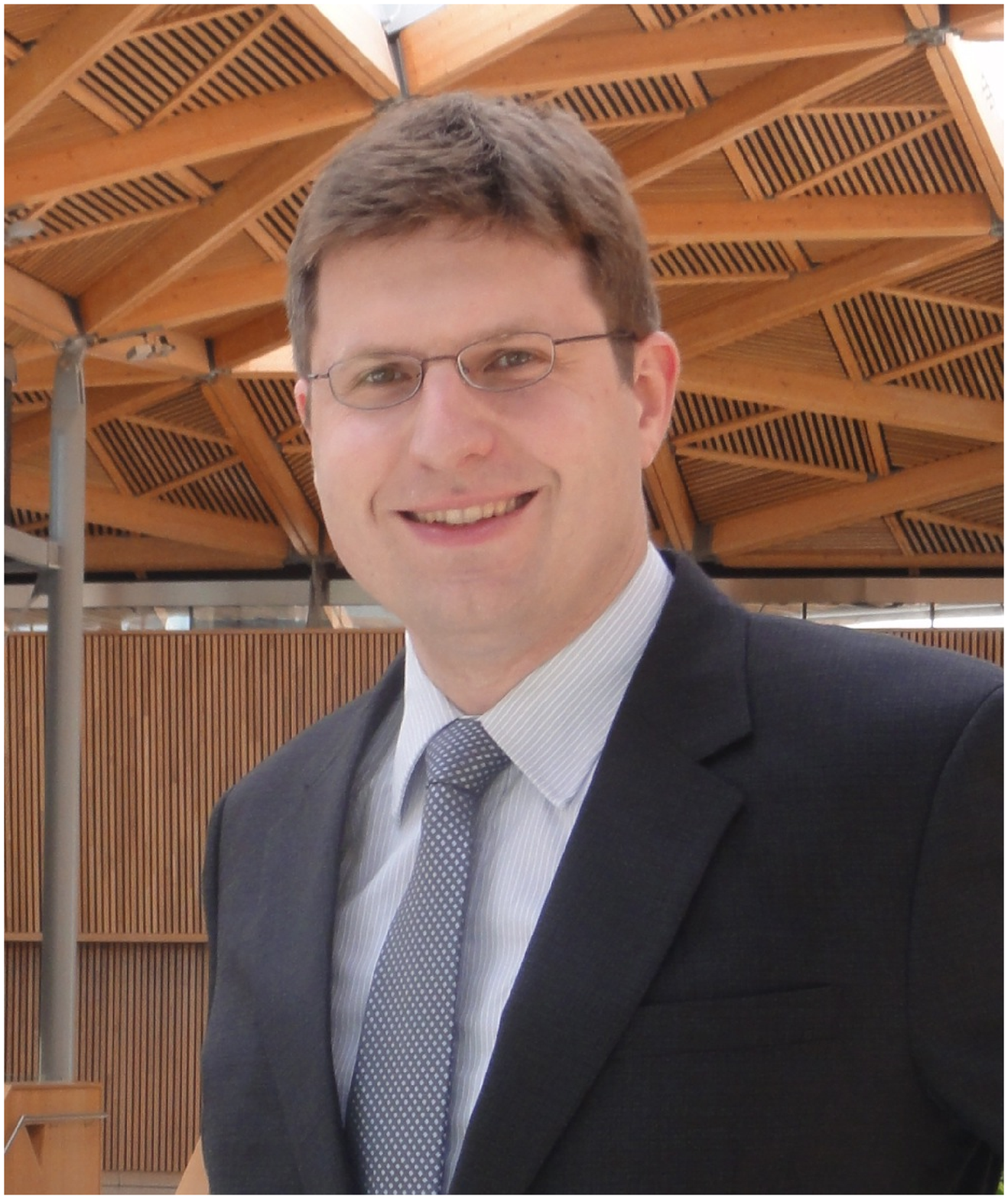}

\vspace{0.5cm}


\end{center}
}}}
\end{center}

\subsection*{Pre-main-sequence multiples as benchmark for disk hydrodynamics}

Understanding how bodies interact with each other and with disk
material holds the key to understanding the architecture of stellar
systems and of planetary systems. While the interactions between point
sources can be described by simple gravity, interactions with disk
material require further knowledge about viscosity and gas+dust
microphysics that need to be included when simulating disk-body
interactions. As a result of our limited knowledge in these areas it
is, for instance, still difficult to estimate the mass of a
gap-opening planet from the morphology of a gap observed in a
protoplanetary disk, or, to derive with certainty whether a gap is
opened by a planet instead of by other processes. Furthermore,
numerical simulations continue to unveil new dynamical processes that
might shape protoplanetary disk structures and affect the planet
populations forming from these disks. One example is {\it disk tearing}
that might occur in the disks around multiple stars whose orbital
angular momentum vectors are misaligned with respect to the rotation
axis of the disk. Based on computer simulations it has been proposed
in 2012 that the resulting gravitational torques could tear the disk
apart and cause rings to separate from the disk and to precess
independently around the central objects (Nixon et al.\ 2012, 2013,
Facchini et al.\ 2013, 2018). In order to test such theories and to
calibrate the fundamental parameters involved in hydrodynamic
simulations, pre-main-sequence (PMS) multiple systems provide us with
a unique laboratory (for general reviews on PMS binaries
and multiples see for instance D\^uchene \& Kraus 2013 and Reipurth et
al. 2014).  
For these systems we are able to directly measure the 3-dimensional
orbits and dynamical masses of the perturbing bodies and can image how
the disk responds to the perturbation.

One system that has the potential to serve as such a ``rosetta stone''
for hydrodynamic studies, is the PMS triple GW\,Orionis. This system
is one of the brightest and best-studied T~Tauri multiple systems. It is
located in the $\lambda$\,Orionis star-forming region (388~pc; Kounkel et al.\ 2017) 
and has an age of $\sim 1$ million years (Calvet et al.\ 2004). With orbital
periods of $\sim 9$\,months and 11\,years, the orbital periods and
separations are just in the right range to enable a full orbit
characterisation, while expecting at the same time strong interactions
between the disk and the stars.

\begin{figure*}[h]
  \begin{center}
  \begin{tabular}{c}
    \vspace{-10mm}
    \begin{minipage}{22cm}
      $\begin{array}{c@{\hspace{10mm}}c@{\hspace{-1mm}}c}
        \multicolumn{3}{c}{\includegraphics[width=0.79\textwidth]{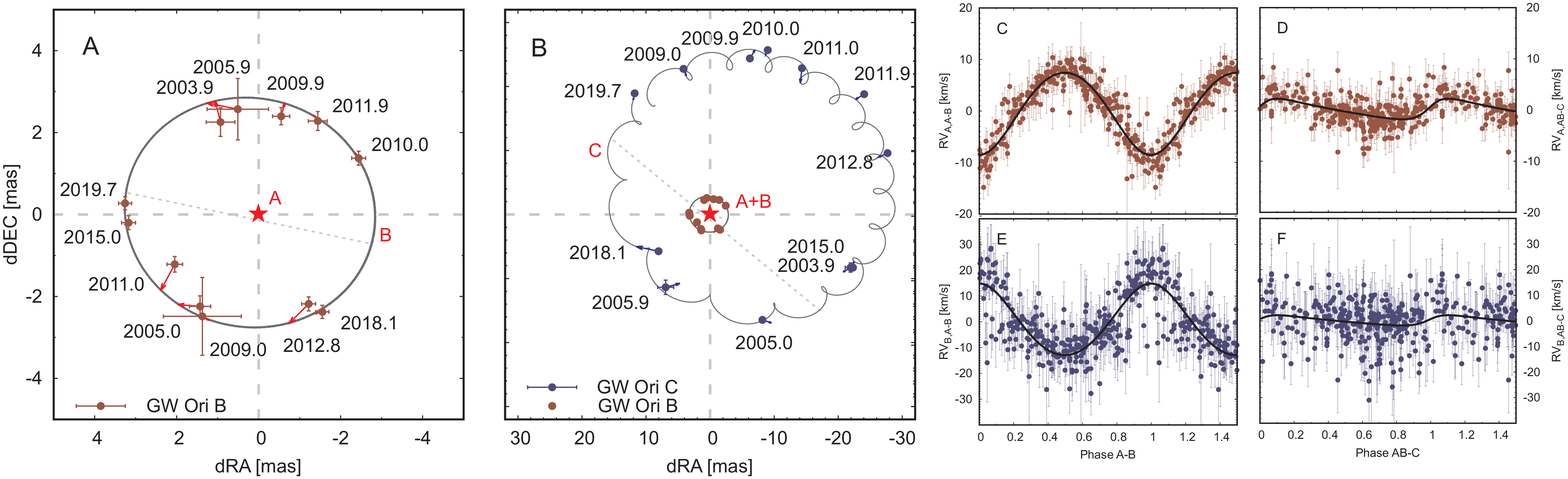}} \\[3mm]
          \includegraphics[width=7cm]{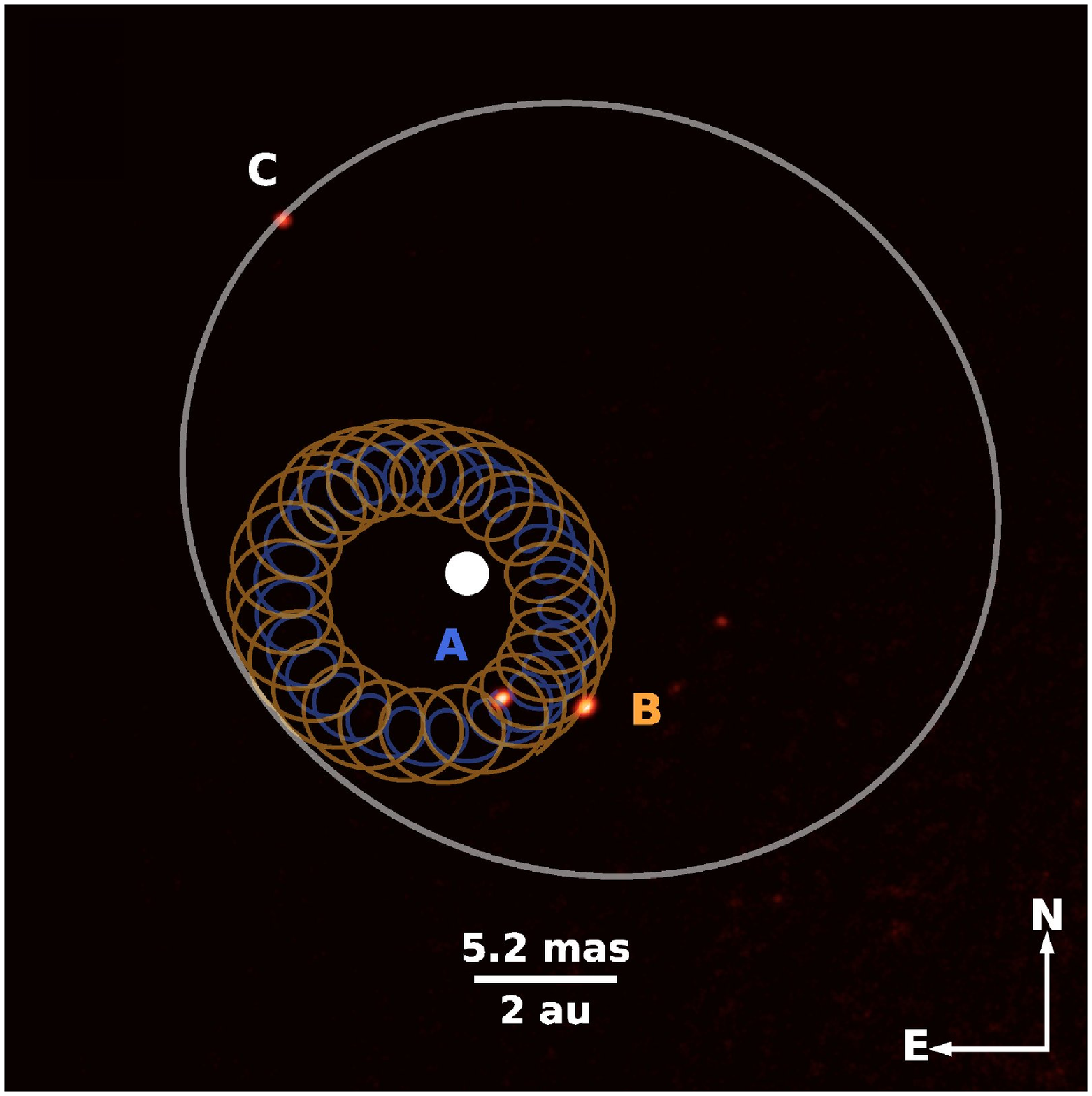} &
          \multicolumn{2}{c}{
            \begin{minipage}{9cm}
              \vspace{-81mm}\caption{
                {\it Top:} GW\,Ori orbit fit to the astrometry of the inner binary (A-B; panel A) and the tertiary (AB-C; panel B), where the primary is in the origin of the coordinate system (Kraus et al.\ 2020a). Panels C-F show the fit to the radial velocity data presented by Czekala et al.\ 2017.\newline
                {\it Left:} Aperture-synthesis image of GW\,Ori, obtained with the MIRC-X instrument and CHARA on 2019 August 27. The white dot marks the location of the center-of-mass of the system, while the blue/brown/white curves give the best-fit orbits derived for the component A/B/C.  From Kraus et al.\ (2020).
              }
            \label{fig:orbit}
          \end{minipage}}
      \end{array}$
    \end{minipage}
  \end{tabular}
  \end{center}
\end{figure*}

\subsection*{The triple star system}

GW\,Ori is long known to be a single-lined spectroscopic binary with a
242\,day period (Mathieu et al.\ 1991). Prato et al.\ (2018) reported
the detection of lines associated with the secondary.  Observations
with the IOTA infrared interferometer resolved the inner binary and
discovered a third component (Berger et al.\ 2011). Building on the work
by Mathieu and coworkers, Czekala et al.\ presented in 2017 an
impressive set of spectra that were obtained over 35\,years with the
Fred L. Whipple Observatory and Oak Ridge Observatory and that provide
radial velocities for the primary and secondary. Using these radial
velocities and the IOTA astrometry, Czekala and colleagues derived
first orbit solutions for all 3 stars. These orbit solutions indicated
a 11.5\,year orbit period for the tertiary and hinted at a signficant
misalignment between the stellar orbits and the disk, although the
small orbital arc covered by the IOTA astrometry resulted in
degeneracies in the orbit fits. Between 2008 and 2019, the VLT
Interferometer and the CHARA array were used to monitor the
astrometric orbit of the inner binary and the tertiary (Kraus et al.\
2020a). The resulting orbits are shown in Figure~\ref{fig:orbit} and
correspond to a near-circular ($e=0.069\pm0.009$),
$241.62\pm0.05$\,day orbit for the inner binary and a
$4216.8\pm4.6$\,day orbit for the tertiary with significant
eccentricity ($e=0.379\pm0.003$), where the mutual inclination between
the orbital planes is $13.9\pm1.1^{\circ}$.

The precise masses of the components in the GW\,Ori system has long
been a matter of debate. Mathieu et al.\ (1995) estimated the mass
using evolutionary tracks, yielding 2.5\,M$_{\odot}$ and
0.5\,M$_{\odot}$ for the primary and secondary, respectively. Other
workers estimated the mass from the H-band flux ratio and derived more
equal mass ratios (e.g.\ 3.2\,M$_{\odot}$ and 2.7\,M$_{\odot}$; Prato
et al.\ 2018). These discrepancies can likely be explained by the
non-negligle \& time-variable contributions from
circumbinary/circumtertiary dust emission biasing the near-infrared
flux ratio (Kraus et al.\ 2020a). The dynamical masses resulting from
the orbit solution are M$_{A}=2.47\pm0.33$, M$_{B}=1.43\pm0.18$, and
M$_{C}=1.36\pm0.28$\,$M_{\odot}$ (Kraus et al.\ 2020a).

\begin{figure*}[t]
\begin{center}
     \includegraphics[width=1.0\textwidth]{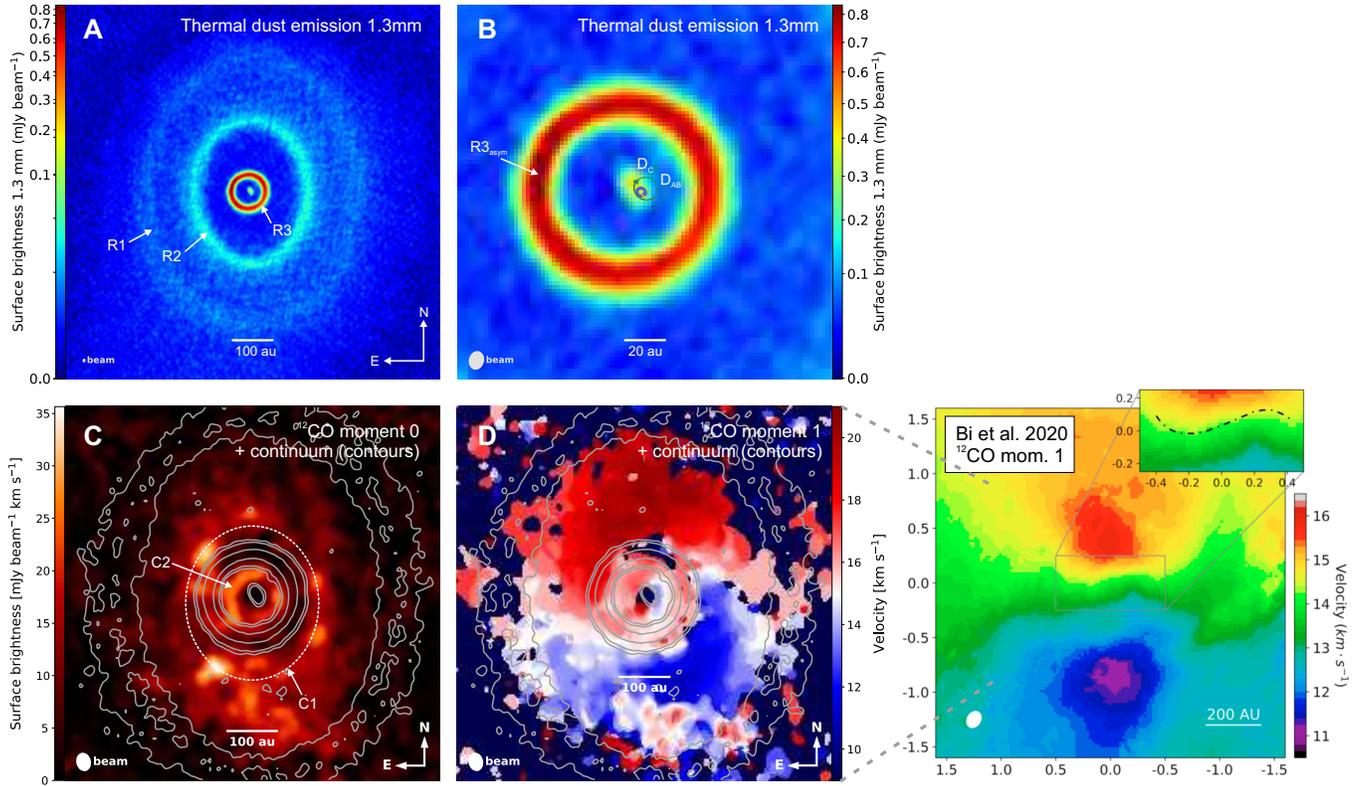}
  \caption{
ALMA observations of GW\,Ori at 0.12'' (right, Bi et al.\ 2020a) and 0.02'' resolution (left, panels A-D, Kraus et al.\ 2020a). The top row shows continuum images, while the bottom row shows $^{12}$CO moment maps.
  }
  \label{fig:ALMA}
\end{center}
\end{figure*}

\subsection*{A highly dynamical environment in the inner few au}

The spectral energy distribution (SED) of GW\,Ori features strong
excess emission from mid-infrared to millimeter wavelengths,
indicating the presence of circumstellar dust. SED modeling suggested
the presence of a circumbinary disk extending from around $1.2$\,au
(Fang et al.\ 2017), 2.1\,au (Artymowicz \& Ludow 1994) or 3.3\,au
(Mathieu et al.\ 1995) outwards. This circumbinary disk was also
resolved spatially, where the VLTI and CHARA visibilities associate
$16\pm 2$\% of the H-band flux with circumbinary material located
$2^{+2.5}_{-0.8}$\,au from the inner binary (Kraus et al.\ 2020a).

To fit the mid-infrared SED, Fang et al.\ (2014, 2017) derived a
dust-depleted gap at $\sim 45$\,au. Further evidence for a truncated
or gapped disk structure comes from the line profile of the CO
fundamental lines, where Najita et al.\ (2003) noted that the line
profile exhibits a narrow+broad emission component and that the line
width increases towards the more energetic transitions. The system
also shows signposts of active accretion, in particular
Br$\gamma$-emission (Folha \& Emerson 2001) and strong UV veiling
($\dot{M}_{\mathrm{acc}}=3 \times 10^{-7}~M_{\odot}$\,yr$^{-1}$,
Calvet et al.\ 2004).

There has been a long debate on the viewing geometry of the system:
Based on measurements of the stellar rotation period ($P=3.25$~days),
the rotation velocity ($v \sin i=43.0 \pm 2.5$~km\,s$^{-1}$) and
estimates of the stellar radius, Bouvier \& Bertout 1989 estimated the
inclination of the system to $\sim 15^{\circ}$, i.e.\ close to
face-on. Using a similar method, but other observational data, Mathieu
et al.\ 1995 obtained an inclination of $30^{\circ}$.  However,
photometric observations also reported Algol-like eclipses (e.g.\
Shevchenko et al.\ 1992, Lamzin et al.\ 1998, Czekala et al.\ 2017)
that have been interpreted as evidence for a nearly edge-on disk
orientation.

Variability on longer time scales has been reported as well, including
dramatic changes in the near-infrared SED on timescales of
$\sim20$\,yrs (Fang et al.\ 2014). This variability might be linked
with a 0.2\,mag-amplitude sinusoidal variation in the V-band light
curve (Czekala et al.\ 2017) that is phased with the orbital period of
the tertiary. The origin of the long-term variability has not been
answered conclusively yet, but might be due to changes in the viewing
geometry or accretion rate on the circumtertiary disk. The circumtertiary
disk has been detected as submillimeter emission in the ALMA 0.02'' images (Fig.~\ref{fig:ALMA}B) 
and as near-infrared excess emission near the location of the tertiary in infrared interferometry data
(Kraus et al.\ 2020a).

\begin{figure*}[t]
  \begin{center}
    \includegraphics[width=0.9\textwidth]{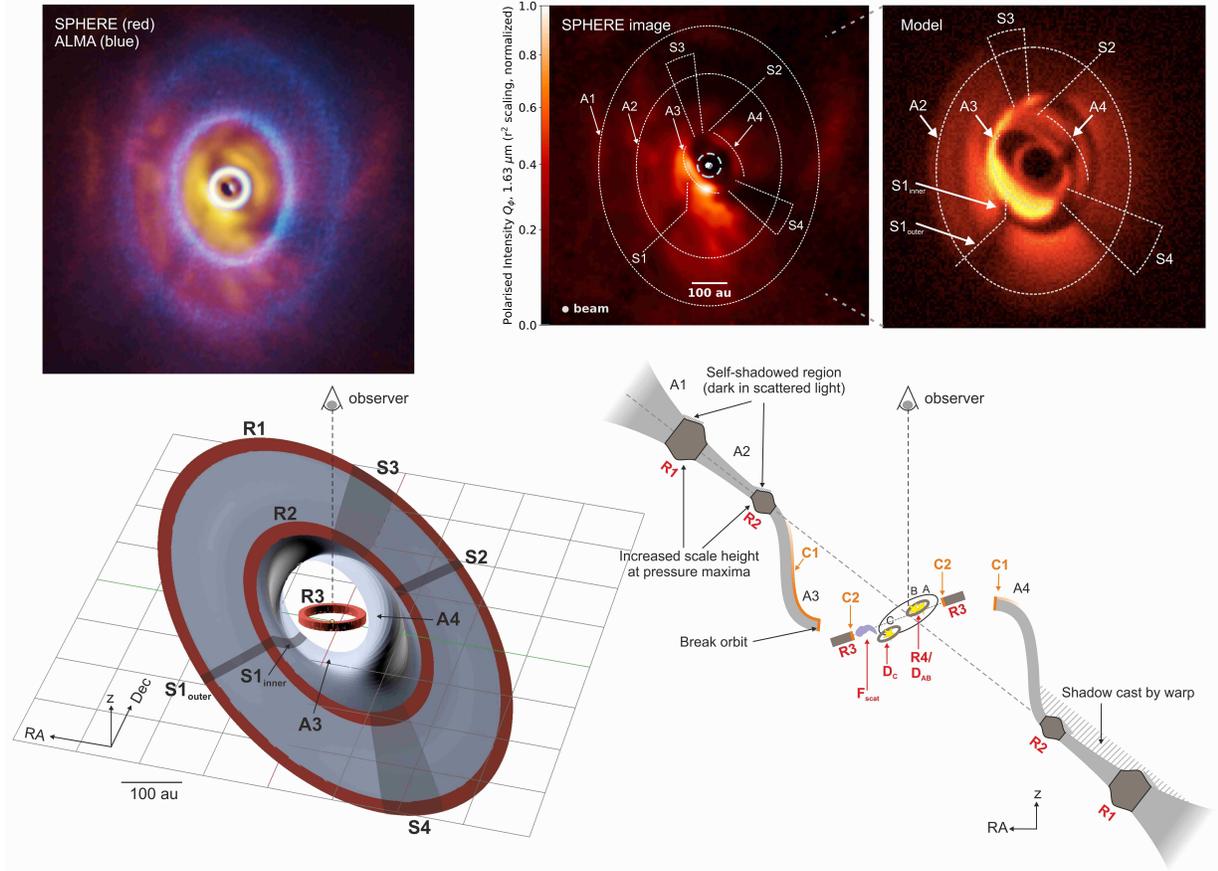}
    \caption{
      Top-left: Overlay of the ALMA continuum image (blue) and the SPHERE scattered light image (red; credit: ESO/Exeter/Kraus et al.). Top-right: SPHERE H-band polarimetric image and model image. Bottom: Sketch of the 3-dimensional disk geometry of GW\,Ori. From Kraus et al.\ 2020a.
    }
    \label{fig:sketch}
  \end{center}
\end{figure*}

\begin{figure*}[h]
  \begin{center}
    \includegraphics[width=0.98\textwidth]{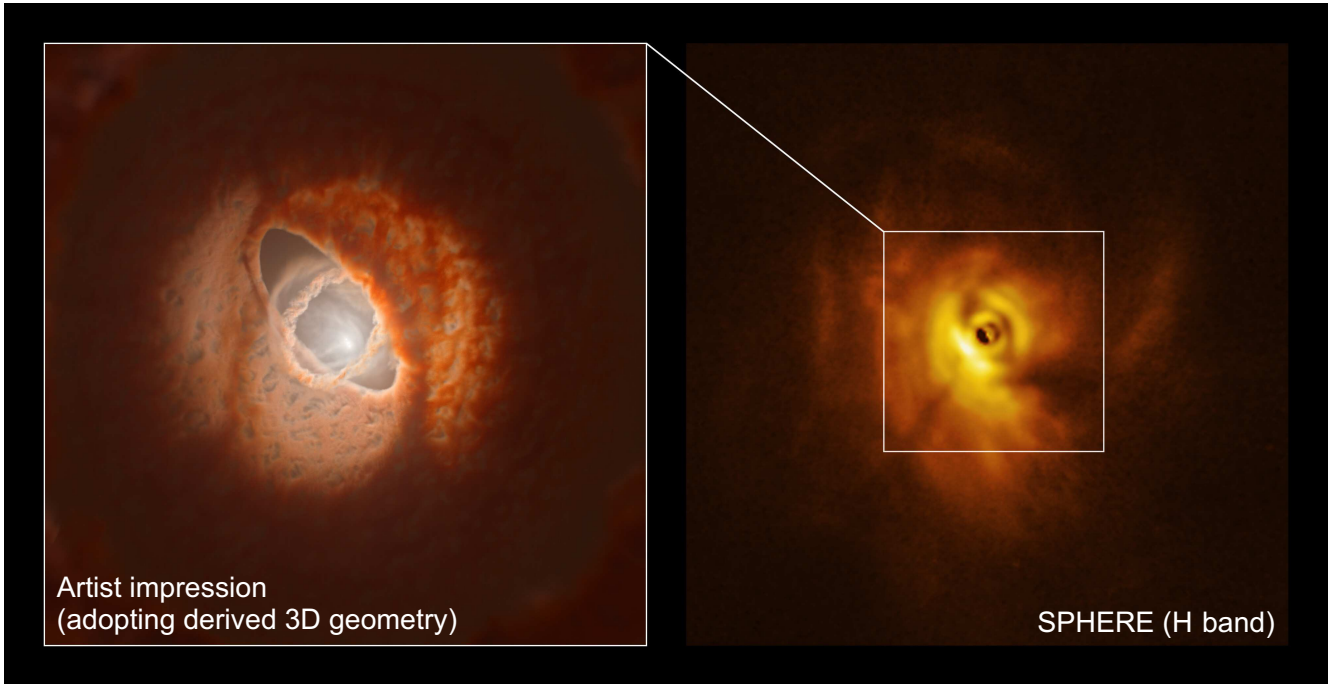}
    \caption{
      Artist impression of the 3-dimensional geometry of the GW\,Ori disk (left) and comparison with the SPHERE image (right). Credit: ESO/Calçada, Exeter/Kraus et al.
    }
    \label{fig:artist}
  \end{center}
\end{figure*}

\begin{figure*}[tbp]
  \begin{center}
    \includegraphics[width=1.0\textwidth]{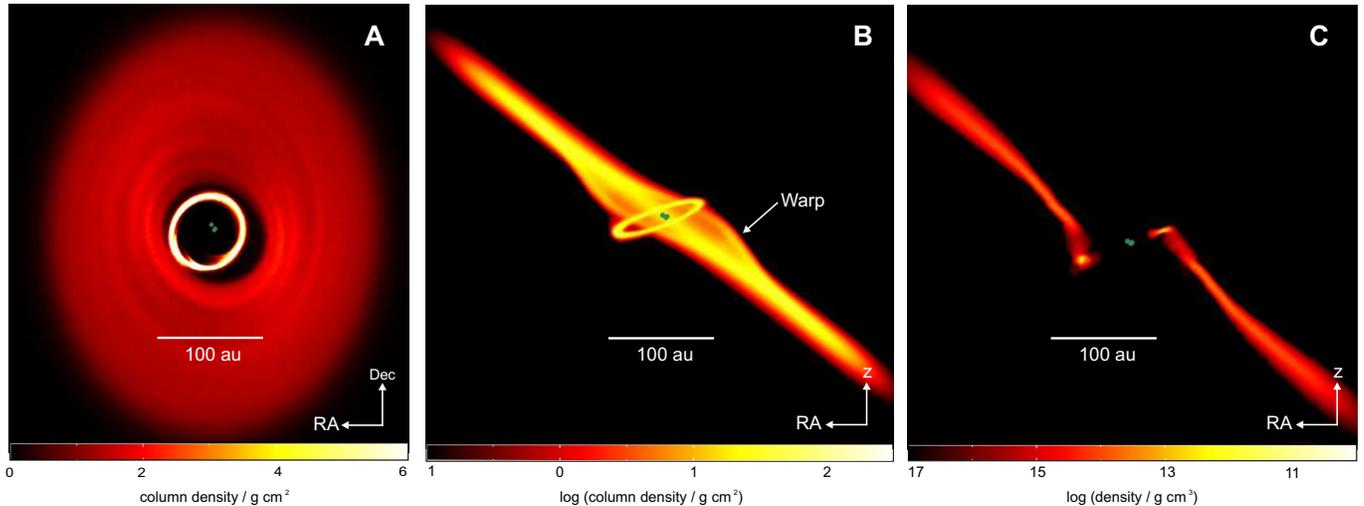}
    \caption{
      SPH model, computed with the sphNG code developed by Matthew Bate and collaborators. The simulation adopts the triple star orbits shown in Fig.~\ref{fig:orbit} and an initial disk orientation that corresponds to the outer ALMA rings R1+R2. The snapshot shows the gas density after 9500\,years. Panel (A) shows the column density along the line-of-sight seen from Earth; in (B) and (C) the z-axis indicates the direction towards the observer.
    }
    \label{fig:SPH}
  \end{center}
\end{figure*}

\begin{figure*}[p]
  \begin{center}
    \includegraphics[width=\textwidth]{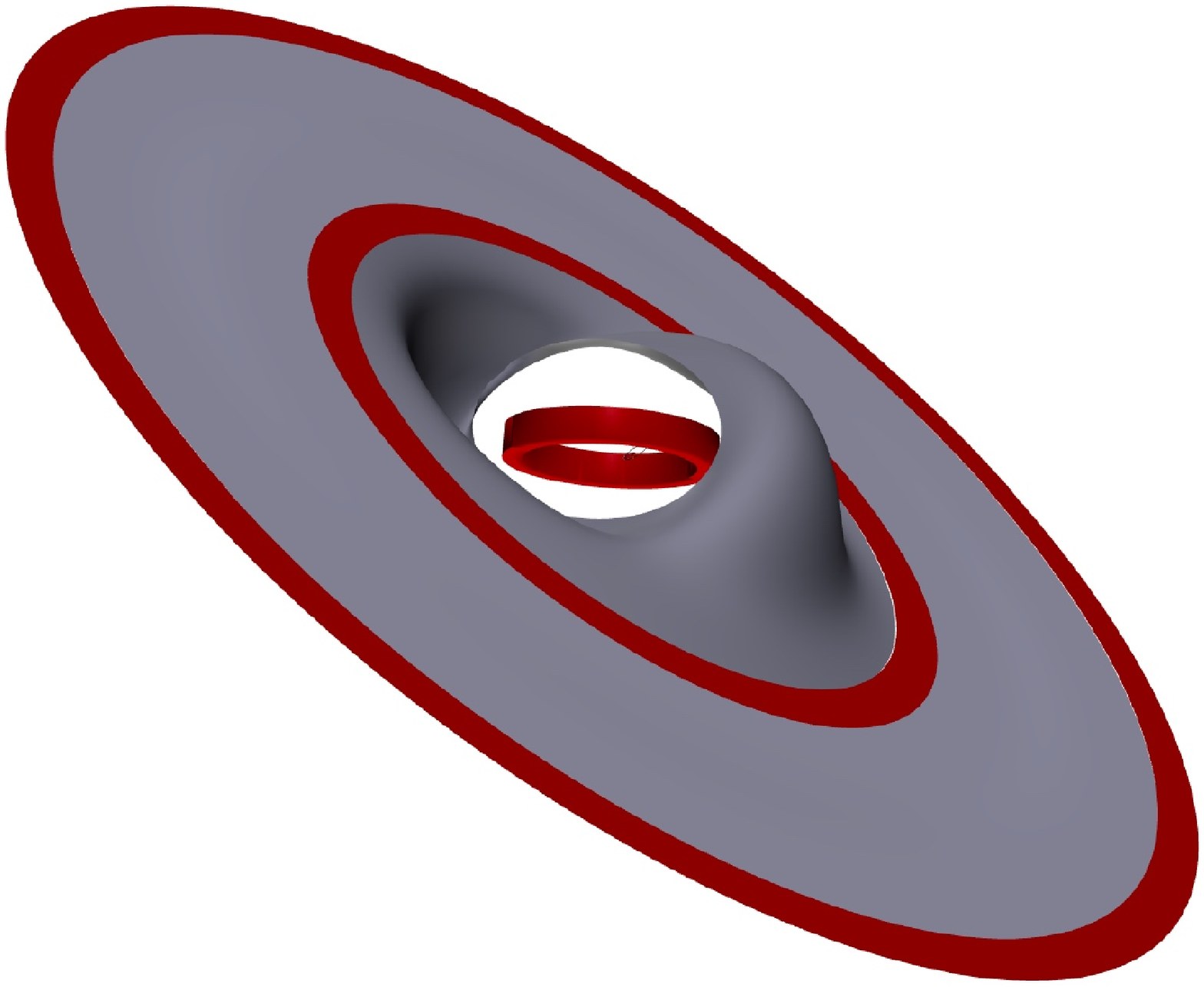}
    \caption{
      The SFN article contains here an interactive 3-dimensional model of the disk geometry in GW\,Ori, as derived in Kraus et al.\ 2020a. Unfortunately, it was not possible to upload the graphics to arxiv -- please retrieve the SFN article from {http://www.ifa.hawaii.edu/users/reipurth/newsletter/newsletter333.pdf} or the 3-dimensional graphics from {https://www.eso.org/public/archives/releases/pdf/eso2014a.pdf} .
    \label{fig:3dmodel}
    }
  \end{center}
\end{figure*}

\subsection*{Misaligned rings}

The intermediate/outer dust disk has been probed with JCMT (Mathieu et
al.\ 1995) and SMA millimeter interferometry, where Fang et al.\ 2017
highlighted the large spatial extent ($\sim 400$\,au) and high mass
(0.12\,M$_{\odot}$) of the disk.  Bi et al.\ 2020 and Kraus et al.\
2020a acquired ALMA data with different baseline configurations and
detected three dust rings.  The two outer rings, R1 and R2, have radii of about
350 and 180\,au and are oriented North-South and seen under intermediate
inclination $37\pm1^{\circ}$ (Fig.~\ref{fig:ALMA}A), representing the
angular moment vector of the cloud that feeds the disk.
The Eastern side of the disk is tilted towards us, as indicated by the
strong forward-scattered light from that side of the disk (Fig.~\ref{fig:sketch}).

The inner submillimeter ring, R3, appears much more circular in the images -- which could
be interpreted as a more face-on viewing geometry. However, from
extrapolating the center of the outer rings (Bi et al.\ 2020) and from
direct imaging (Kraus et al.\ 2020a) it is evident that the inner ring
is not centered on the position of the stars
(Fig.~\ref{fig:ALMA}B). This can be explained best if the ring is
intrinsically eccentric but seen under significant inclination and
appears near-circular in projection.  The following additional
information can be used to derive the 3-dimensional shape and
orientation of the ring: 
\begin{itemize} 

\item[(a)] {\bf Gas kinematics}: $^{12}$CO moment 1 maps show that the
rotation axis of the outer disk is oriented roughly in East-West
direction (position angle $90^{\circ}$; e.g.\ Fang et al.\ 2017,
Czekala et al.\ 2017) with a 'twist' in the velocity field in the
inner 0.2'' (Bi et al.\ 2020; Fig.~\ref{fig:ALMA}, right). The twist
might follow a spiral-arm pattern, with the position angle of the
rotation axis changing to $180^{\circ}$ at 100\,au and
$\sim210^{\circ}$ at $\sim 30$\,au (Kraus et al.\ 2020a,
Fig.~\ref{fig:ALMA}D).

\item[(b)] {\bf Warm gas at the inner surface of the ring}: The
$^{12}$CO moment 0 map shows that the CO surface brightness is low at
the location of the ring R3, which can be explained with the high optical
depth and low gas temperature within the ring.  However, there is
strong CO emission near the inner edge of ring R3 on the Eastern side
(labeled C1 in Fig.~\ref{fig:ALMA}C), indicating that the Eastern side
of the ring is farther away from the observer and that we see warm gas
at the illuminated inner surface of the ring (Fig.~\ref{fig:sketch},
bottom-right; Kraus et al.\ 2020a).

\item[(c)] {\bf Shadows cast by the ring:} SPHERE scattered light
imagery (Figs.\ 3 and 4, top) exhibits several shadows, including
narrow shadows in south-east and north-west direction (S1 and S2;
Fig.~\ref{fig:sketch}, top) and broader shadows extending in north and
south-west direction (S3 and S4). Remarkably, shadow S1 does not
follow a straight line but changes direction at $\sim 100$\,au
separation (Fig.~\ref{fig:sketch}, top-right), indicating that the
shadow falls onto a curved surface in the inner 100\,au.  Simultaneous
modeling of the shadow morphology and of the ALMA continuum geometry
yields an eccentricity $e=0.3$ and a semi-major axis of 43 au radius
for ring R3 and that the ring is seen under inclination of
$155^{\circ}$ (Fig.~\ref{fig:sketch}; Kraus et al.\ 
2020a). \end{itemize}

\subsection*{Broken \& warped disk geometry}

Shadows have been observed in several protoplanetary disks, but
GW\,Ori is (to my knowledge) the first case where the ring casting the
shadow has been spatially resolved.
This enables tight constraints on the shape and 3-dimensional orientation 
of the misaligned ring as well as the curvature of the warped disk surface 
inside of the middle ring R2 ($r \lesssim 182$\,au).  
The scattered-light morphology shows a strong
East-West asymmetry, where the bright Eastern arc A3 and the fainter
Western arc A4 form together an apparent ellipse with semi-major axis
of 90\,au and high eccentricity ($e=0.65$; Fig.~\ref{fig:artist}).  
Kraus et al.\ (2020a) identifies this ellipse as the point
where the disk breaks due to the gravitational torque from the central
triple system. The bright arc A3 constitutes the side of the warped
disk surface that is facing away from Earth and that appears bright in
scattered light due to the direct illumination from the stars. Arc A4
corresponds to the side facing towards us, where we see only the
self-shadowed outer side of the warped surface (Fig.~\ref{fig:sketch}).
The shadows from the misaligned ring are cast onto this warped surface and 
appear as shadows S1 and S2, while the broad shadows S3 and S4 correspond to the
regions where the break orbit crosses the plane of the outer disk,
which coincides with the direction in the warp with the highest radial
column density.

\subsection*{Origin of the disk misalignments}

To determine the origin of the extreme disk misalignments observed in
GW\,Ori, two teams recently presented smoothed particle hydrodynamic
simulations. Bi et al.\ conducted SPH simulations using the 'phantom'
code and concluded that the gravitational torque of the stars alone
cannot explain the observed large misalignment between the dust ring. 
Instead, they propose that an undiscovered companion located
between the inner and middle ring that might have broken the disk and
induced the misalignments.

The 'sphNG' simulation presented in Kraus et al.\ (2020a), on the
other hand, shows the disk tearing effect, where the gravitational
torque of the three stars tears the disk apart into distinct rings
that precess independently around the central objects. After letting
the dust distribution evolve for a few thousand years, a ring breaks
out of the disk plane, whose radius ($\sim 40$\,au), eccentricity
($e\sim 0.2$), and extreme misalignment agree well with the observed
properties of ring R3. Also, the disk breaks and a warp forms just
beyond the break radius, whose dimension, geometry, and low column
density agree reasonably well with the properties of the warp derived
from the GW\,Ori observations (Fig.~\ref{fig:SPH}).

Both simulations adopt similar Shakura-Sunyaev viscosities
($\alpha_{\mathrm{SS}}=0.008-0.013$ for Bi et al.; 0.01-0.02 for Kraus
et al.). Therefore, it appears more likely that the different outcomes
might be related to the setup of the simulation. There are differences
concerning the number of stars included in the simulation (2 stars in
Bi et al. simulation; 3 stars in Kraus et al. simulation) and the
orbit solution that is adopted for the simulation (Czekala et al.\
2017 solution and Kraus et al.\ 2020a solution, respectively).

\subsection*{Outlook}

Over the last few decades several exciting pre-main-sequence multiple
systems have been found and extensively studied, including GG\,Tau,
HD142527, HD98800, and T\,Tauri.  GW\,Ori stands out with respect to
the tight constraints on the full 3-dimensional orbits for {\it all} components 
in the system, the dynamical masses, and our knowledge on the 
3-dimensional geometry of the strongly distorted disk
(for a visualization of the deduced disk geometry \& orbits, see the
interactive 3-dimensional model in Fig.~6).
Due to this unique information, the system could serve as a
valuable benchmark for calibrating hydrodynamic models and fundamental
parameters under well-defined conditions. This could provide the
validation \& refinement that is needed before applying the models to
the much less-well-constrained planet formation case, where the masses
and orbits of the gap-opening bodies are not known in general.

The disk-tearing effect that we might witness in GW\,Ori {\it in
action}, constitutes an important new mechanism for moving disk
material onto highly oblique or retrograde orbits, even at very wide
separations from the star. At the same time, the observed torn ring
seems to be sufficiently massive and might be sufficiently stable for
planet formation to occur, potentially giving rise to an
yet-undiscovered population of circum-multiple planets on highly
oblique, long-period orbits.

An important open question concerns the origin of the outermost ring
in GW\,Ori. The high submillimeter brightness of the inner and middle ring
can likely be explained by disk tearing and dust filtration processes
near the disk warp region. It is unclear whether the outer-most ring
can also be explained by the dynamical interplay between the central
triple system and the disk, or whether dust trapping near a
planet-induced density gap might be required to explain the high
submillimeter surface brightness in this region.

But what caused the misalignment between the disk and the orbits in
the first place?  Possibilities include turbulent disk fragmentation
(Offner et al.\ 2010), perturbation by other stars in a stellar
cluster (Clarke \& Pringle 1993), the capture of disk material during
a stellar flyby (Clarke \& Pringle 1991), or the infall of material
with a different angular momentum vector from that of the gas that
formed the stars initially (Bate et al.\ 2010, Bate 2018). To answer
this question, statistical information will be of essence, both on the
disk-orbit misalignment in pre-main-sequence multiples and on the
orbital architecture and spin-orbit alignment in main-sequence systems.
Obtaining such constraints on a large sample of stars is a
key science objective for the proposed VLTI instrument BIFROST (Kraus
et al.\ 2019, 2020b) and could offer important new insights
on both the star- and planet-formation processes.

\footnotesize

{\bf Acknowledgement:}

I would like to thank my collaborators on our recent paper on GW\,Ori
and the members of the team around Jiaqing Bi for delightful
discussions on this fascinating object.  Also I acknowledge support
from an ERC Starting Grant (Grant Agreement No.\ 639889).\newline

{\bf References:}

Artymowicz \& Ludow 1994, ApJ 421, 651\\
Bate, M.R., et al.\ 2010, MNRAS 401, 1505\\
Bate, M.R., 2018, MNRAS 475, 5618\\
Berger, J.-P., et al.\ 2011, A\&A 529, L1\\
Bi, J., et al.\ 2020, ApJL 895, L18\\
Bouvier, J. \& Bertout, C. 1989, A\&A 211, 99\\
Calvet, N., et al.\ 2004, AJ 128, 1294\\
Clarke, C.J., \& Pringle, J.E.\ 1991, MNRAS 249, 588\\
Clarke, C.J., \& Pringle, J.E.\ 1993, MNRAS 261, 190\\
Czekala, I., et al.\ 2017, ApJ 851, 132\\
D\^uchene, G. \& Kraus, A. 2013, ARAA, 51, 269\\
Facchini, S., et al.\ 2013, MNRAS 433, 2142\\
Facchini, S., et al.\ 2018, MNRAS 473, 4459\\
Fang, M., et al.\ 2017, A\&A 603, 132\\
Fang, M., et al.\ 2014, A\&A 570, 118\\
Folha, D.F.M. \& Emerson, J.P., 2001, A\&A 365, 90\\
Kraus, S., et al.\ 2019, zenodo.3356286\\
Kraus, S., et al.\ 2020a, Science 369, 6508\\
Kraus, S., et al.\ 2020b, ApJ 897, 8\\
Kounkel, M., et al.\ 2017, AJ 834, 142\\
Lamzin, S., et al.\ 1998, Ap\&SS, 261, 167\\
Mathieu, R.D., et al.\ 1991, AJ, 101, 2184\\
Mathieu, R.D., et al.\ 1995, AJ, 109, 2655\\
Najita, J., et al.\ 2003, ApJ 589, 931\\
Nixon, C., et al.\ 2012, ApJ 757, L24\\
Nixon, C., et al.\ 2013, MNRAS 434, 1946\\
Offner, S., et al.\ 2010, ApJ 725, 1485\\
Prato, L., et al.\ 2018, ApJ 852, 38\\
Reipurth, B. et al. 2014, Protostars and Planets V, p.\ 267\\
Shevchenko, V.S., et al.\ 1992, IBVS 3746, 1

\end{document}